\documentstyle[aps,prl,twocolumn,epsf,floats]{revtex}

\begin{document}
\draft

\wideabs{

\title{Hund's Rule for Monopole Harmonics, or
       Why the Composite Fermion Picture Works}

\author{
   Arkadiusz W\'ojs$^*$ and John J. Quinn}
\address{
   The University of Tennessee, Knoxville, Tennessee, 37996}

\maketitle

\begin{abstract}
The success of the mean field composite Fermion (MFCF) picture in 
predicting the lowest energy band of angular momentum multiplets in 
fractional quantum Hall systems cannot be found in a cancellation 
between the Coulomb and Chern--Simons interactions beyond the mean 
field, due to their totally different energy scales. 
We show that the MFCF approximation can be regarded as a kind of 
semi-empirical Hund's rule for monopole harmonics. 
The plausibility of the rule is easily established, but rigorous 
proof relies on comparison with detailed numerical calculations.
\end{abstract}

}

It is well known that the mean field composite Fermion (MFCF) picture 
\cite{jain} correctly predicts the low lying band of angular momentum
multiplets of a fractional quantum Hall (FQH) system by simply noting 
that when $N$ electrons are converted to $N$ composite Fermions (CF's), 
the angular momentum of the lowest shell goes from $l_0$ to $l_0^*=l_0
-p_0(N-1)$, where $p_0$ is an integer.
A very fundamental question, which is not understood, is ``Why does the 
MFCF picture work so well in describing not just the Jain sequence of
incompressible ground states, but also of the low lying band of multiplets
for any value of the filling factor $\nu$?''
The answer cannot lie in the cancellation between the Coulomb and
Chern--Simons interactions among the fluctuations because these 
interactions are associated with different energy scales. 
In this note we demonstrate that the predictions of the MFCF picture
can be thought of as a Hund's rule governing monopole harmonics, which
selects a low lying angular momentum subset of the allowed $L$ multiplets 
associated with low values of the Coulomb repulsion.
The plausibility of the rule is established by proving that:
(i) the pseudopotential describing the Coulomb repulsion for monopole
harmonics decreases rapidly  as the pair angular momentum $L_{12}$ 
decreases from its maximum value $L_{12}^{\rm MAX}=2l_0-1$;
(ii) multiplets with lower values of the total angular momentum $L$ 
have, on the average, lower values of $\left<\right.\hat{L}_{ij}^2\left.
\right>$, the expectation value of the pair angular momentum $\hat{L}_{ij}
=\hat{l}_i+\hat{l}_j$;
(iii) low angular momentum values $L$ for which many independent multiplets
occur are more likely to have some low lying multiplets than neighboring
$L$ values with few multiplets; and
(iv) relatively higher multiplicities tend to reoccur at the same $L$ 
values for different values of $l_0$.

For $N$ electrons on a Haldane sphere \cite{haldane} (containing at the
center a magnetic monopole of charge $2S\,hc/e$), the single particle
states fall into angular momentum shells with $l_n=S+n$, $n=0,1,\dots$
The CF transformation attaches to each electron a flux tube of strength 
$2p_0$ flux quanta oriented opposite to the original magnetic field.
If the added flux is treated in a mean field approximation, the resulting
effective magnetic field is $B^*=B-2p_0\,(hc/e)\,n_s$ ($n_s$ is the number 
of electrons per unit area). 
An effective CF filling factor, ${\nu_0^*}^{-1}=\nu_0^{-1}-2p_0$,
and an effective monopole strength seen by one CF, $2S^*=2S-2p_0(N-1)$, 
can also be defined.
$|S^*|$ plays the role of the angular momentum of the lowest CF shell
\cite{chen}.
States belonging to the Jain sequence occur when $\nu_0^*$ is an integer.
For such integral CF fillings, the ground state is a Laughlin 
\cite{laughlin} incompressible liquid state with angular momentum $L=0$.
If $\nu_0^*$ is not an integer, a partially occupied CF shell will contain 
$n_{\rm QP}$ quasiparticles (QP's).
In the MFCF picture these states form a degenerate band of angular 
momentum multiplets with energy $n_{\rm QP}\varepsilon_{\rm QP}$, where 
$\varepsilon_{\rm QP}$ is the energy of a single QP.
The degeneracy results from the neglect of QP--QP interactions in the 
MFCF approximation \cite{sitko1}.

The single particle states for an electron on a Haldane sphere are called
monopole harmonics \cite{wuyang} and denoted by $\left|l_n,m\right>$, 
where $-l_n\le m\le l_n$. 
The single particle energies depend only on $S$ and $n$, and for the FQH 
effect, only the lowest shell with $n=0$, which is completely spin 
polarized, need be considered.
The object of numerical studies is to diagonalize the electron--electron
interaction within the subspace of the ${2S+1\choose N}$ many particle 
states of the lowest shell.
The numerical calculations become difficult when the number of electrons 
$N$ exceeds 10.
The calculations give the eigenvalues $E$ as a function of the total
angular momentum $L$, and the numerical results always show one or more
$L$ multiplets forming a low energy sector (or low energy band).

The problem of $N$ Fermions in a shell of angular momentum $l$ is very
familiar from atomic physics \cite{cowan}.
In this note we investigate the analogy between the problems of $N$ 
electrons in the lowest angular momentum shell of a Haldane sphere and 
$N$ electrons in an atomic shell of the same angular momentum.
{\em First}, because $2S$ is an integer, the monopole harmonics can have 
integral or half-integral orbital angular momentum.
The spherical harmonics have $S=0$, so $l$ must be an integer.
For FQH systems (i.e. $\nu<1$) we are interested in the lowest angular 
momentum shell with $l_0=S$.
{\em Second}, for FQH systems, calculations with $N$ values up to 10 and 
$l$ values up to 27/2 have been performed \cite{fano}, while in atomic 
system $l$ values up to 3 ($f$-states) and $N$ values up to 7 are usually 
the maximum values studied.
{\em Third}, the Zeeman splitting is large compared to the Coulomb 
interaction, so only totally spin polarized states of FQH systems need 
be considered.
The total spin is always equal to ${1\over2}N$, and the total (spin {\em 
plus} orbital) angular momentum is simply the sum of $L$ and ${1\over2}N$.
Thus, only the second Hund's rule is of interest; it states that the largest 
allowable value of $L$ (consistent with maximum possible spin) will be the 
ground state.
This is certainly not the case for FQH systems.
Many Laughlin incompressible states at $L=0$ are ground states, and states
containing 1, 2, 3, \dots QP's always have allowed $L$ values that are much 
smaller than $L_{\rm MAX}={1\over2}N(2l-N+1)$.
What causes this difference?

\begin{figure}[t]
\epsfxsize=3.4in
\epsffile{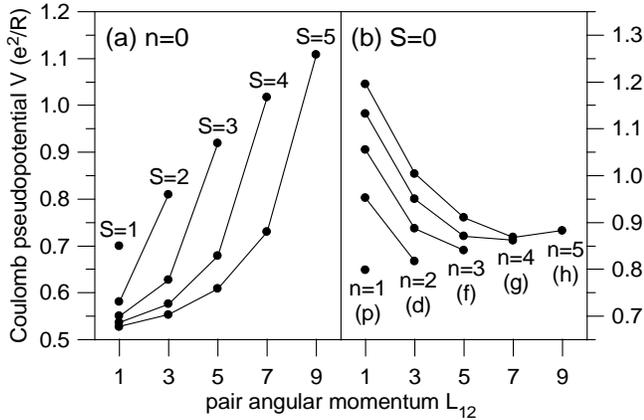}
\caption{
   The pseudopotential for the pair of electrons of total angular 
   momentum $L_{12}$ as a function of $l=S+n$.
   Energy is measured in units of $e^2/R$, where $R$ is the radius 
   of the sphere.
   (a) monopole harmonics, $n=0$;
   (b) spherical harmonics, $S=0$, calculated for a radial wave 
   function which localizes the electrons at radius $R$.
   A plot similar to (a) [$V(L_{12})$ vs. $L_{12}^{\rm MAX}-L_{12}$]
   for values of $2S$ up to 25 is given in \protect\cite{wojs}}
\label{fig1}
\end{figure}

In Fig.~\ref{fig1} we display the Coulomb pseudopotential for a pair of 
electrons in single particle angular momentum states with $l=1$ through 5, 
as a function of the pair angular momentum ${\bf L}_{12}={\bf l}_1+{\bf l}_2$.
For monopole harmonics ($l=S$, $n=0$) $V(L_{12})$ increases with increasing
$L_{12}$.
For atomic shells (spherical harmonics) just the opposite occurs -- the 
repulsion decreases with increasing $L_{12}$ (for the $h$-shell and higher, 
$V(L_{12})$ begins to increase beyond some relatively large value of 
$L_{12}$, but this is never of concern in atomic physics).
The function $V(L_{12})$ is obtained by diagonalizing the Coulomb 
interaction within the space of antisymmetric pair wave functions.
The different behavior of monopole harmonics is due to the Lorentz force
caused by the electron--electron repulsion in the presence of the uniform
magnetic field.
A pair of electrons which are close together have large total angular
momentum and large repulsion.

It is useful to write an antisymmetric wave function $\left|l^N,L\alpha
\right>$ for $N$ electrons each with angular momentum $l$ that are combined
to give a total angular momentum $L$ as \cite{cowan}
\begin{equation}
\label{eq1}
   \left|l^N,L\alpha\right>=
   \sum_{L_{12}}\sum_{L'\alpha'} 
   G_{L\alpha,L'\alpha'}(L_{12})
   \left|l^2,L_{12};l^{N-2},L'\alpha';L\right>.
\end{equation}
Here $G_{L\alpha,L'\alpha'}(L_{12})$ is called the coefficient of 
fractional grandparentage.
In Eq.~(\ref{eq1}), $\left|l^2,L_{12};l^{N-2},L'\alpha';L\right>$ is 
a state of angular momentum $L$.
It is antisymmetric under permutation of particles 1 and 2, which have
pair angular momentum $L_{12}$, and under permutation of particles 3, 4,
\dots, $N$, which have angular momentum $L'$.
The label $\alpha$ (or $\alpha'$) distinguishes independent orthogonal
states with the same angular momentum $L$ (or $L'$).

A very useful operator identity
\begin{equation}
\label{eq2}
   \hat{L}^2 + N(N-2)\;\hat{l}^2 = \sum_{\rm pairs} \hat{L}_{ij}^2
\end{equation}
is straightforward to prove.
Here $\hat{L}=\sum_i\hat{l}_i$ and $\hat{L}_{ij}=\hat{l}_i+\hat{l}_j$.
Taking the expectation value of Eq.~(\ref{eq2}) in the state $\left|l^N,L
\alpha\right>$ gives
\begin{eqnarray}
\label{eq3}
 & & \left<l^N\!\!\!,L\alpha\right|
      \sum_{\rm pairs}\hat{L}_{ij}^2
   \left|l^N\!\!\!,L\alpha\right> 
 =L(L+1)+N(N-2)\,l(l+1)
\nonumber\\
 & & ={1\over2}N(N-1) \sum_{L_{12}}
   {\cal G}_{L\alpha}(L_{12})\; L_{12}(L_{12}+1).
\end{eqnarray}
In this equation ${\cal G}_{L\alpha}(L_{12})=\sum_{L'\alpha'}\left|
G_{L\alpha,L'\alpha'}(L_{12})\right|^2$.

From the orthonormality of the functions $\left|l^N,L\alpha\right>$
it is apparent that $\sum_{L_{12}}{\cal G}_{L\alpha}(L_{12})=1$, and
\begin{equation}
\label{eq4}
   \sum_{L_{12}} \sum_{L'\alpha'}
   G^*_{L\alpha,L'\alpha'}(L_{12})\; G_{L\beta,L'\alpha'}(L_{12})
   =\delta_{\alpha\beta}.
\end{equation}
The energy of the state $\left|l^N,L\alpha\right>$ is given by
\begin{equation}
\label{eq5}
   E_\alpha(L)=\sum_{L_{12}}{\cal G}_{L\alpha}(L_{12})\;V(L_{12}).
\end{equation}
It is noteworthy that the expectation value of $\sum_{\rm pairs}
\hat{L}_{ij}^2$ is independent of which multiplet $\alpha$ is being 
considered.
In view of Eqs.~(\ref{eq3}) and (\ref{eq5}), it is not surprising that 
in atomic physics, where $V(L_{12})$ decreases rapidly with $L_{12}$, 
Hund's second rule holds.
For states with $L=L_{\rm MAX}$ only a single multiplet ever appears, 
and it has the highest value of the average pair angular momentum.
Despite this strong indication that, in atomic systems, the state with 
the largest allowed value of $L$ has the lowest energy, Hund's rule is 
considered an empirical rule, that can be rigorously justified only by 
numerical calculations.

For the case of monopole harmonics, $V(L_{12})$ decreases very rapidly 
as $L_{12}$ decreases from its maximum value $L_{12}^{\rm MAX}=2l_0-1$.
Therefore, low energy multiplets must somehow be able to avoid having 
large grandparentage in states with large values of $L_{12}$.
In a previous paper we have demonstrated analytically that this is true
for three electron systems \cite{wojs}.
For the monopole harmonics the general trend is to have $N_L^{-1}
\sum_\alpha E_\alpha(L)$, the average $E(L)$ for all multiplets with 
angular momentum $L$, increase with increasing $L$.
However, when the single particle angular momentum, $l$, increases beyond
some value for an $N$ particle system, several multiplets of the same $L$ 
begin to appear.
In Tab.~1 we present as an example, the number of independent multiplets of 
angular momentum $L$ as a function of $2S$ for a system of eight electrons.
\begin{table}
\caption{
   The number of independent multiplets at angular momentum $L$ 
   for eight electrons as a function of $2S$ for $0\le2S\le22$.
   Only $L$ values up to 8 are included in the table}
\begin{tabular}{r|rrrrrrrrr}
  $_{2S}\mbox{}^{L}$
     &  0&  1&  2&  3&  4&  5&  6&  7&  8\\ \hline
    0&   &   &  1&   &   &   &   &   &   \\ 
    1&  1&   &  1&   &  1&   &   &   &   \\ 
    2&  1&   &   &   &   &   &   &   &   \\ 
    3&  1&   &  1&   &  1&   &   &   &   \\ 
    4&  1&   &  1&  1&  1&   &  1&   &   \\ 
    5&  1&   &  1&   &  1&   &  1&   &   \\ 
    6&   &   &   &   &  1&   &   &   &   \\ 
    7&  1&   &   &   &   &   &   &   &   \\ 
    8&   &   &   &   &  1&   &   &   &   \\ 
    9&  1&   &  1&   &  1&   &  1&   &  1\\ 
   10&  1&   &  1&  1&  2&  1&  2&  1&  1\\ 
   11&  2&   &  3&  1&  4&  2&  4&  2&  4\\ 
   12&  2&  1&  4&  3&  6&  5&  7&  5&  7\\ 
   13&  4&  1&  7&  5& 11&  7& 13&  9& 13\\ 
   14&  4&  3& 10&  9& 16& 14& 19& 17& 21\\ 
   15&  7&  4& 16& 13& 25& 21& 31& 26& 35\\ 
   16&  8&  8& 21& 22& 35& 33& 45& 42& 51\\ 
   17& 12& 10& 32& 30& 51& 48& 66& 61& 77\\ 
   18& 13& 17& 42& 45& 69& 70& 91& 90&108\\ 
   19& 20& 22& 58& 61& 96& 95&128&124&152\\ 
   20& 22& 33& 75& 85&126&133&169&173&205\\ 
   21& 31& 42&101&111&168&175&227&230&277\\
   22& 36& 59&126&150&215&233&294&307&360
\end{tabular}
\label{tab1}
\end{table}
The values of $2S$ go from zero to twenty two; the values of $L$ are shown
up to eight.
If the pseudopotential were given by $\tilde{V}(L_{12})=A+B\,L_{12}(L_{12}
+1)$, all of the different multiplets with the same value of $L$ would
be degenerate because of Eqs.~(\ref{eq3})--(\ref{eq5}), and $L_{\rm MIN}$, 
the smallest allowed $L$ multiplet, would be the ground state.
The difference between $\tilde{V}(L_{12})$ and the actual pseudopotential 
$V(L_{12})$ leads to a lifting of this degeneracy (different multiplets 
repel one another).
The splittings caused by $V(L_{12})-\tilde{V}(L_{12})$ can become large
when $N_L$, the number of times the multiplet $L$ occurs, is large.
In this case, a state with $L$ larger than $L_{\rm MIN}$ can become
the ground state since the actual values of $E_\alpha(L)$ depend on how
the values of ${\cal G}_{L\alpha}(L_{12})$ are distributed, not just on
the average value of $\hat{L}_{12}^2$ for that value of $L$.
For example, the lowest energy multiplet with $L=4$ at $2S=20$ is lower 
in energy than the multiplets at $L=0$, 1, 2, and 3.
The same is true of the lowest energy multiplet with $L=4$ at $2S=22$.
Knowing which multiplet is the ground state or which multiplets form the 
``low energy sector'' without performing detailed numerical calculations
is a considerably more difficult task than it was for spherical harmonics.
It is very likely however, that the highest $L$ value corresponds to the 
highest energy.

As might be expected, when the angular momentum $l_0$ of the lowest 
electron shell is replaced by $l_0^*=l-p_0(N-1)$, the possible values of 
the resulting total angular momentum $L^*$ are less than or equal to 
a value $L^*_{\rm MAX}$, that is always small compared to $L_{\rm MAX}$.
For example, if $2S\ge3N-3$, $L^*_{\rm MAX}={1\over2}N(2S-3N-3)$; if $3N
-3\ge2S\ge{5\over2}N-4$, $L^*_{\rm MAX}=(2S-{5\over2}N+4)(3N-3-2S)$; etc.
At filling factors corresponding to states in the Jain sequence $L^*_{\rm 
MAX}=0$.
For states containing one or more QP's, a number of different $L^*$ values 
less than or equal to $L^*_{\rm MAX}$ can occur.
From the numerical calculations it has been observed \cite{jain,haldane} 
that the subset of allowed $L^*$ multiplets obtained by placing $N$ CF's 
into the lowest angular momentum shells forms the low energy sector of the 
spectrum of the original electron system.
This is plausible because:
(i) the allowed values of $L^*$ are always small compared to the original
$L_{\rm MAX}$ and therefore have a small expectation value of the pair
angular momentum $\hat{L}_{ij}$, and
(ii) the low values of $L$ which occur a relatively large number of times 
tend to form the low energy band of values $L^*$.
For the eight electron system with a given value of $2S$, the allowed 
$L^*$ values are those appearing in the row with $2S^*=2S-14$.
The table of multiplicities depends only on $|2S|$, so if $2S-14$ is 
negative, it is simply replaced by its magnitude.
Because of this, the $\nu=2/3$ state occurs at $2S=12$, and the $\nu=2/5$ 
state occurs at $2S=16$.

\begin{figure}[t]
\epsfxsize=3.4in
\epsffile{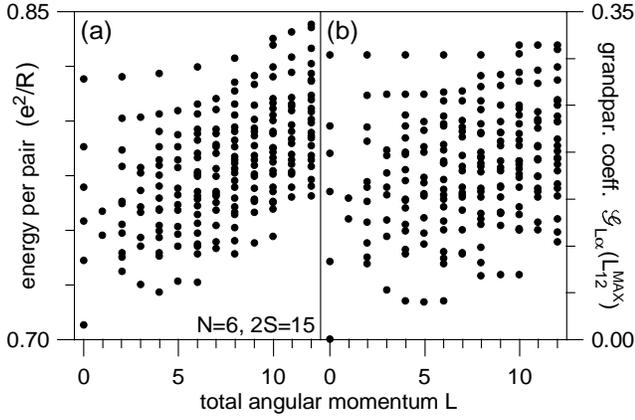}
\caption{
   Energy spectrum (a) and ${\cal G}_{L\alpha}(L_{12})$ (b) as 
   a function of $L$ for a system of six electrons at $2S=15$ 
   (Laughlin $\nu=1/3$ state)}
\label{fig2}
\end{figure}

We have evaluated ${\cal G}_{L\alpha}(L_{12})$ for values of $N\le8$ and 
for many different values of $2S$.
In Fig.~\ref{fig2} we compare the energy spectrum (a) for a six electron
system at $2S=15$ (the Laughlin $\nu=1/3$ state) with the coefficient
(b) ${\cal G}_{L\alpha}(L_{12}^{\rm MAX})$ (where $L_{12}^{\rm MAX}=14$),
the coefficient associated with the maximum Coulomb repulsion.
The similarity of the two figures makes it clear that a model 
pseudopotential with only $V(L_{12}^{\rm MAX})$ non-vanishing reproduces
the main features of the low energy spectrum.

\begin{figure}[t]
\epsfxsize=3.4in
\epsffile{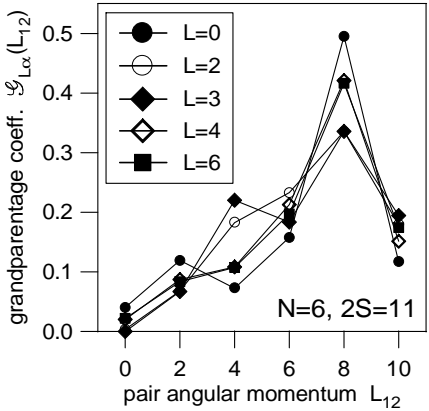}
\caption{
   ${\cal G}_{L\alpha}(L_{12})$ vs. $L_{12}$ of the lowest energy 
   multiplets at $L=0$, 2, 3, 4, and 6 for $N=6$ and $2S=11$
   (corresponding to the $\nu=2/5$ state).
   Note that ${\cal G}_{L\alpha}(L_{12}^{\rm MAX})$ is smaller
   for $L=0$ than for any of the other states}
\label{fig3}
\end{figure}

In Fig.~\ref{fig3} we plot ${\cal G}_{L\alpha}(L_{12})$ vs. $L_{12}$ for
a six electron system with $2S=11$ for the lowest multiplets having $L=0$,
2, 3, 4, and 6.
The $L=0$ state is the Jain incompressible ground state at $\nu=2/5$.
The other states contain a single QP pair. 
Notice that ${\cal G}_{L\alpha}(L_{12}^{\rm MAX})$ is smaller for the 
$L=0$ ground state than it is for the neighboring states.
Because $V(L_{12}^{\rm MAX})$ is so large, this coefficient dominates in
the determination of the energy.

Two additional points are worth emphasizing \cite{wojs}.
{\em First}, the CF hierarchy containing all odd denominator fractions
\cite{sitko2} is obtained by reapplying the MFCF transformation to residual 
QP's in a partially filled CF shell.
However, in order for the MFCF approach to be valid, the QP--QP interaction
has to be similar to the Coulomb pseudopotential, falling rapidly from its
maximum value $V(L_{12}^{\rm MAX})$ with decreasing $L_{12}$.
This is not true \cite{sitko2} for all QP--QP interactions, suggesting why 
the states of the Jain sequence are the most stable incompressible liquid 
ground states.
{\em Second}, states containing a single quasihole (e.g. the lowest energy 
state for $2S=3(N-1)+1$) have ${\cal G}_{L\alpha}(L_{12}^{\rm MAX})=0$ just 
as the neighboring Laughlin state (at $2S=3(N-1)$) does.
However, the single quasielectron state (at $2S=3(N-1)-1$) cannot have 
${\cal G}_{L\alpha}(L_{12}^{\rm MAX})=0$.
Because $V(L_{12}^{\rm MAX})$ is so large, $\varepsilon_{\rm QE}$ is much 
larger than $\varepsilon_{\rm QH}$.

We have demonstrated that the MFCF picture selects a low angular momentum
subset of the allowed set of $L$ multiplets for $N$ electrons on a Haldane
sphere.
We make the hypothesis that this set of low angular momentum multiplets 
forms the low energy sector of the spectrum, and offer arguments that
support this hypothesis.
These arguments make our hypothesis plausible, but (as with Hund's rule for 
atomic spectra) the proof lies in comparison with detailed calculations.
For every case we have studied ($N\le8$ and many different values of $2S$),
the probability ${\cal G}_{L\alpha}(L_{12})$ for the large repulsive part
of the Coulomb interaction is found to be smaller for the $L$ values
predicted by the MFCF picture than for neighboring states, verifying that
the MFCF picture acts as a Hund's rule for monopole harmonics.

The authors gratefully acknowledge support from the Division of Materials 
Sciences -- Basic Energy Research Program of the U.S. Department of Energy,
and thank D. C. Marinescu, K. S. Yi, and P. Sitko for useful discussions.
A. W. acknowledges support from the KBN Grant No. PB674/P03/96/10.

$^*$ On leave from the Institute of Physics, Wroc\l aw University of 
Technology, Poland.

\end{document}